# CREATIVITY TRAINING FOR FUTURE ENGINEERS: PRELIMINARY RESULTS FROM AN EDUCATIVE EXPERIENCE

## Sophie Morin[1], Jean-Marc Robert[2], Liane Gabora[3]

[1]École Polytechnique de Montréal (CANADA)
[2] École Polytechnique de Montréal (CANADA)
[3]University of British Columbia (CANADA)

## Abstract

Due in part to the increased pace of cultural and environmental change, as well as increased competition due to globalization, innovation is become one of the primary concerns of the 21st century. We will present an academic course designed to develop cognitive abilities related to creativity within an engineering education context, based on a conceptual framework rooted in cognitive sciences. The course was held at École Polytechnique de Montréal (ÉPM), a world renowned engineering school and a pillar in Canada's engineering community. The course was offered twice in the 2014-2015 academic year and more than 30 students from the graduate and undergraduate programs participated. The course incorporated ten pedagogical strategies, including serious games, an observation book, individual and group projects, etc., that were expected to facilitate the development of cognitive abilities related to creativity such as encoding, and associative/analytical thinking. The CEDA (Creative Engineering Design Assessment) test was used to measure the students' creativity at the beginning and at the end of the course. Field notes were taken after each of the 15 three-hour sessions to qualitatively document the educative intervention along the semester and students gave anonymous written feedback after completing the last session. Quantitative and qualitative results suggest that an increase in creativity is possible to obtain with a course designed to development cognitive abilities related to creativity. Also, students appreciated the course, found it relevant, and made important, meaningful learnings regarding the creative process, its cognitive mechanism and the approaches available to increase it.

# 1 CONTEXT

Innovation has become one of the 21st century's most important matters [1]. There appears to be an emerging worldwide consensus that the future of profitable businesses and organizations depend strongly on their innovation capacity. Experts in innovation are talking about the "open innovation era": "*Firms that can harness outside ideas to advance their own businesses while leveraging their internal ideas outside their current operations will likely thrive in this new era of open innovation.*" [2: 41] The United Nations published a report that explained and promoted the value of a creative economy [3].

Around the world, public and private organizations are making innovation the cornerstone of their strategic economic development [4, 5]. Moreover, as of 2007 in the US, creative industries represented 11% of the GDP (Gross Domestic Product) [6].

The European Union named 2009 as the "The European Year of Creativity and Innovation" [7: 1]. The same year, three Canadian funding agencies (Social Sciences and Humanities Research Council of Canada, Natural Sciences and Engineering Research Council of Canada, Canadian Institutes of Health) published the Government's recommendations for future funding and stated that "*Research excellence and creativity are now understood as key assets for national success and international competitiveness.*" [8: 1]

In the US in 2004, the National Academy of Engineering published its vision of the 2020 engineer. They presented several important "soft skills" engineers should master such as leadership, communication, and creativity [7]. Since 2010, the Canadian Engineering Accreditation Board (CEAB) has encouraged engineering faculties to develop 12 engineers' qualities within their curricula. Apart professionalism, engineering knowledge and communication, CEAB includes conception (systems conception, processes conception, engineering tools and instruments conception), a term that has the same etymology as creativity, which call upon creativity training.

Innovation, design, engineering, and competitiveness are undoubtedly closely related. As professionals, engineers must consistently come up with innovative ideas to keep their organization competitive. The driver of such innovative capacity is creativity. Engineering schools are therefore striving to provide students with not just the tools of their trade but also with creativity fostering tools. Although creativity is related to problem solving, which is covered in engineering curricula, a different perspective must be taken to find appropriate answers to innovative behaviors. Creativity research provides a wealth of theories, concepts, models, tools and results that can be used in creativity training. Cognitive processes related to divergent thinking are believed to be essential components of the training required and should be incorporated into curricula to improve student's creative potential [9, 10, 11].

The time has come for educators to adapt and remodel their programs in ways that foster creative skills [1, 11, 12, 13]. Though steps have been taken (for example, the establishment of university degrees specialized in creativity, introduction to creativity principles in science faculties, etc.) the lack of research and structure in the theoretical frameworks, course designs, training techniques and instructional media, , leaves teachers and educational program designers with very little empirical evidence on which to base effective and efficient programs [14, 15, 16].

## 2   TEACHING AND LEARNING CREATIVITY

We identified five ways creativity training is added in university programs; there are specific degrees in creativity (graduate, postgraduate, masters, etc.); creativity courses or workshops not specific to a program; creativity consultants; and written materials suggested in project courses (examples are shown in Table 1). They are based on diverse philosophies, approaches, activities, etc. We suspect that numerous engineering courses include "time for creativity" directly in their syllabi, requiring students to show creativity in their work. In engineering education, Project Based Learning (PBL) became more popular to help students apply learned concepts and allow them to "think outside the box" or "be creative" [17, 18].

Table 1 – Creativity training programs examples

| Training Categories | Location | School |
|---|---|---|
| Courses in Programs (1 or 2 courses dedicated to creativity) | France | *École Nationale Supérieure en Génie des Systèmes et de l'Innovation* (ENSGSI)<br>–   Cursus Engineering; Profile Conception-Innovation |
| | USA | *Oakland University*<br>–   Course – Creativity & Innovation |
| Degrees (graduate, postgraduate, masters,etc.) | USA | *International Center for Studies in Creativity* (Buffalo State University)<br>–   Master program/Graduate Certificate |
| | | *Edward de Bono Institute for the Design and development of Thinking* (University of Malta)<br>–   Master in Creativity and Innovation |
| | | *Texas A&M University*<br>–   Minor in Creative Studies |
| | Spain | *Universitad Fernando Pessoa de Compostela*<br>–   International Masters in Creativity, International Doctorate in Creativity, Masters in Creativity and Innovation |
| Trainings | UK | *The Thinking Business*<br>–   Creative Thinking Training Course |
| | Canada | *Polytechnique Montréal*<br>–   "Creativity, Yes you can" (Mandatory for PhD students, 12 hours) |
| | Canada | *Hautes Études Commerciales* (HEC Montréal)<br>–   MOSAIC (Creativity Management) |
| Consultants | Canada | *Idea Connection* (British Colombia) |

| | | *Zins Beauchesne and Associates* (Quebec) |
|---|---|---|
| | France | *Good Morning Creativity* (Paris) |

This brings up some important considerations. Does encouraging students to be creative efficiently and sufficiently makes them more creative? Perhaps. Would more structured instructions be more effective? Perhaps.

Scott's work demonstrates how much leeway is taken in every creativity training program aspect (course design, techniques, media, etc). She and her colleagues did a meta-analysis of 70 different creativity trainings [16, 17]. As we know of, although their two articles are 10 years old, they are the most recent available sources and contain the most detailed information available on training programs [16, 17]. Using a rigorous five criteria methodology to select eligible programs, they analyzed pedagogical issues of 70 programs and their effectiveness on learning creativity. Their conclusions are useful but modest: "(1) training should be based on a sound, valid, conception of the cognitive activities underlying creative efforts, (2) training should be lengthy and relatively challenging with various discrete cognitive skills, and associated heuristics, (3) articulation of these principles should be followed by illustrations of their application using material based on "real-world" cases, (4) presentation of this material should be followed by a series of exercises appropriate to the domain at hand".[14: 383]

In these articles, the authors present many findings concerning pedagogical elements of the programs. One is the listing of different types of media chosen by various authors to teach creativity (e.g.: lectures, video or audio, computer assisted, individualized coaching, case based, programmed instruction, discussion, cooperative learning, etc.). This list illustrates the diversity of the media selected. The article also adds the effectiveness associated with each media but unfortunately, it is not related to any other variables (topic, length, exercises, etc.). Furthermore, the authors suggest 11 training types created by clustering individual elements identified in the studied programs. These clusters are described as training in: analogy, analytical, open idea production, interactive idea production, creative process, imagery, creative/critical thinking, situated idea production, structured idea production, conceptual combination, and computer assisted idea production.

This constitutes a valid foundation in better understanding what kind of training works, but not enough to build a strong, reliable training program. Their conclusions leave too many unanswered design concerns to be used as the only guide prepare creative training programs. Moreover, their results are 10 years old and considering the topic's popularity emergence in the last decade, we can only speculate on how it is different now.

The vision is that before using creativity methods, techniques (*e.g.*, brainstorming, mind mapping, SCAMPER, etc.), there is a need to understand and develop and practise elementary cognitive abilities (like a musician practising scales before playing a piece, or a dancer doing *barre* exercises). To create a new course aiming at the improvement of future engineer's creativity, we decided to establish the course on a conceptual framework based on cognitive science and more precisely an intelligence model presented in the next section.

# 3   CONCEPTUAL FRAMEWORK

Creativity has been defined as "the ability to produce work that is novel (i.e., original, unexpected), high in quality, and appropriate (i.e., useful, meets task constraints)"[19]. Alternatively, Wilkenfeld and Ward [20] define creativity as "The activation and recombination in a new way of previous knowledge elements in order to generate new properties based on the previous ones". Combining elements of both these definitions, we define creativity as the skill to produce a new, original, and useful artefact or behavioral outcome that responds to an identified need, by recombining concepts in a new way. Creativity must be considered as much a process as a result.

The need to study creativity, to understand its mechanisms and develop related cognitive skills has been recognized  by the scientific communities more than 60 years ago [4, 13, 21]. Considering the numerous training programs offered in schools and by consultants, on websites and in books, the demand and the interest towards creativity trainings in all domains is  obvious [22]. Unexpectedly, data accessibility about the numerous existing programs is a challenge and information about their content, structure, training methods and activities, training personnel qualifications, and impact is scarce.

We believe that a conceptual framework is essential to elaboration and implement a course on creativity. Numerous models of the creative process exist [11, 23, 24]. For example, Sawyer presents eight different models before suggesting his own [9]. These models present the creativity process as phases such as preparation, incubation, illumination and verification, missing to discuss important specific cognitive abilities linked to the creative process. To define that feature, we constructed a model from the integration of elements of the Cattel-Horn-Carroll model (CHC) [25] and other cognitive abilities now known to be related to creativity [26, 27, 28, 29].

This model suggests different levels of cognitive abilities linked to creativity. Concept combination and cognitive processes (associative thinking, analytical thinking, encoding, and potentiality) are associated with more fundamental cognitive abilities (*e.g.*: associative memory, ideational fluency, expressional fluency, figural flexibility, meaningful memory, etc.). These cognitive abilities can then be targeted by specific pedagogical strategies. The understanding of the creative process stops being only procedural (steps to solve a problem) and becomes deeper, at a cognitive level.

There seems to be an agreement amongst the education community that creativity can be taught and improved [4] but there is a need to better understand the teaching and learning strategies that can foster creativity.

# 4   METHODOLOGY

An intervention study was planned with a multi-method quantitative/qualitative descriptive design. A 45-hour course was created, named (IND8108) "Creativity in applied sciences and engineering". It is offered to all EPM engineering students, undergraduates and graduates. Contents and educative activities are listed in Table 1. All activities aim to the development of cognitive abilities that are believed to support creativity. At the end of the semester, students of all engineering programs (graduates and undergraduates) will have improved cognitive abilities to better accomplish a creative process. They will understand how to access different thinking paths and respond to the instruction "think outside the box". It allows one to acquire a mix of knowledge and competence in creativity. The Table 2 presents the general descriptions of the pedagogical activities developed for the new course: Creativity in applied sciences and engineering.

Table 2 – IND 8108 – Content and pedagogical strategies

| Pedagogical Activities | | *Objectives* |
|---|---|---|
| **In class activities** | Warm-up exercises<br>• ½ h at the beginning of every class, different versions of the same exercises, individual and group | Cognitive abilities (associative and analytical thinking, encoding, potentiality) |
| | Conference<br>• 1 ½ h, invited speaker, presentation + exercises | Improvisation, expand lateral knowledge, draw analogies with engineering |
| | "MouseTrap" Project<br>• Group project, students make their own designs, 2 ½ classes, build a "*Rube Goldberg* machine" | Experience a live group construction project |
| | Lectures/Discussions<br>• Discussions about factors influencing creativity, myths and truths, definitions, etc. | Expand knowledge on creativity, meta-cognition |
| | Creativity approaches<br>• Small groups (2-3), research a given approach and present it in class (theory/exercise), 15 presentations | Acquire and share knowledge about creativity stimulations approaches |
| **Projects** | Artistic or personal Project<br>• Individual project, students choose their subject, first half of semester, presentation to group | Expand knowledge in different domains, try new things, meta-cognition in report |
| | Engineering project<br>• Group project, students choose a problematic they want to tackle, second half of semester, presentation to group | Apply new cognitive abilities, apply creativity approaches, team work |
| **Personal Work** | Log book<br>• Individual task, spread on all semester, evaluate a few times during the semester, | Being a better observer, more critical towards what you see, improve problem finding skill |
| | students note observations, ideas, problems, solutions, etc. | |
| | Creative behaviors/examples (forum)<br>• Ongoing through the semester, posting on the course's website | Share creative behaviors between colleagues (pictures, websites, articles, etc.) |
| | Scientific Texts (forums)<br>• Individual reading, group discussion on course website with guided questions (4 articles during semester) | Expand knowledge on creativity and scientific work |

The course started conservatively as an option in the "*Innovation technologique*" orientation. However, a more ambitious training program could also be considered. For these cognitive abilities to stay anchored and useable by the learners, continuous practise is important [4, 9]. Supported by comments made by students, we suggest two creativity trainings (*e.g.*, autumn of $1^{st}$ and $3^{rd}$ year) completed with the implication of creativity educators in the integrative projects (spring of every year). That will allow great opportunities to master the cognitive abilities and apply them to engineering projects.

# 5 RESULTS

## 5.1 Intervention (course) and participants

The course was offered on two semesters (Fall 2014 and Winter 2015) as an elective course to graduate and undergraduate students. Table 3 describes main characteristics of the students who attended the course.

Table 3 – Sample description

| Criteria | Semester – Fall 2014 | Semester – Winter 2015 |
|---|---|---|
| Number of student | 9 | 22 |
| Genders<br>(Female ; Male) | 4 ; 5 | 10 ; 12 |
| Education<br>(Grads ; Undergrads) | 9 ; 0 | 7 ; 15 |
| Age<br>(20-29; 30-39; 40-49) | 7 ; 1 ; 1 | 18 ; 4 ; 1 |

At the beginning of the first class, students were asked to introduce themselves and to explain their motivation to get in course. All were intrigued by such a "different course at EPM". Many believed that it is important to be "creative engineers".

## 5.2 Quantitative results

The CEDA (Creative Engineering Design Assessment) is an instrument that has been developed by Charyton [30]. Especially made for engineering students, it gives a score based on a judging process. This test is inspired by the *Purdue Creative Test (PCT)* [31], well known in the creativity field [32]. Unlike other creativity tests [33, 34], the CEDA takes 5 different aspects of the creative process into account. Table 4 presents these aspects and how the CEDA assesses them. Usually, creativity tests measure exclusively divergent thinking skills, which are sometimes confused with creativity itself. However, we believe that creativity is more than divergent thinking and that its assessment should cover more aspects of the individual creative process. Especially for engineers, who should be able to satisfy constraint, identify new opportunities (problem finding) and solve problems. By using the test on the first and the last class, we expected to find an improvement in the individual scores.

Table 4 – Creative process measured by CEDA [35]

| Creative Process | Elements measured |
|---|---|
| Divergent Thinking | 2 to 4 different solutions to each problem |
| Convergent Thinking | One solution to the given problem |
| Constraint Satisfaction | Shapes used and materials added within the parameters of design |
| Problem Finding | Identifying other uses for their design |
| Problem Solving | Solving the given problem with a novel or novel designs |

### 5.2.1 Using the CEDA

When completing the CEDA test, participants have to come up with a total of six different designs. Three problems are given, each asking for two different designs or solutions (make sound, communicate and travel). For each problem, general shapes are proposed to construct the designs (cube, cylinder, sphere and pyramid) but any other objects can be added. Students have to complete five categories: sketch, description, materials used, other problems solved and users. Four criteria are used to assess the tests: fluidity, flexibility, originality and usefulness.

The test lasts 30 minutes and it gives a numerical score allowing for the comparison between students and, most importantly, with themselves if the test is repeated. The author specifies that the higher the score obtained the more creative a person is. However, there are no indications as what are the levels of performance. What is a "good" or "bad" or "average" score? No results have ever been published with actual test results.

To study the impact of the course' design, we used the CEDA in a pre-post experiment. On the 2nd and the 14th class, we asked the students to do the same test. No indications were given the first time to suspect there would be a second time. As suggested by Charyton, the same test was given at the end of the course (13 weeks later), in the same class conditions. The synthetized results are presented in the table below (Table 5).

Table 5 – Pre-Post experiment quantitative results

| | Pre-Post Gap | | | | | Pre-Post Gap | | | |
|---|---|---|---|---|---|---|---|---|---|
| | *Criteria* | | | | | *Criteria* | | | |
| | *Fluidity* | *Flexibility* | *Originality* | *Usefulness* | | *Fluidity* | *Flexibility* | *Originality* | *Usefulness* |
| S01 | 11 | 14 | -1,5 | -1 | S17 | 9 | 8 | 0 | 8 |
| S02 | -13 | -13 | 4 | 0,5 | S18 | -3 | 1 | 4 | -3 |
| S03 | 3 | -2 | 0 | 1,5 | S19 | -2 | -5 | -10 | 3 |
| S04 | 4 | 1 | 0 | -0,5 | S20 | 11 | 9 | -4 | 0 |
| S05 | 2 | 3 | 13 | 5 | S21 | -14 | -14 | -2 | -3 |
| S06 | 9 | 8 | 9,5 | 9 | S22 | 8 | 8 | 1,5 | 2 |
| S07 | 19 | 12 | 0 | 5 | S23 | 20 | 22 | 8,25 | 2,5 |
| S08 | 3 | 0 | -16,5 | -4,5 | S24 | 32 | 30 | -3,5 | -5,75 |
| S09 | -2 | 1 | 0 | 8 | S25 | 1 | -1 | -0,75 | 0,75 |
| S10 | 3 | 6 | 0,5 | 3,5 | S26 | 6 | 5 | 1,25 | 0,5 |
| S11 | 43 | 32 | 5 | 3,5 | S27 | -4 | -4 | 16 | 0,75 |
| S12 | 5 | 7 | 3,5 | 7 | S28 | 31 | 29 | 14 | 9,25 |
| S13 | 3 | 1 | -3,5 | -1,5 | S29 | 2 | 0 | 5 | -2,75 |
| S14 | 11 | 11 | 5 | 6,5 | S30 | 22 | 20 | 5,25 | 0,75 |
| S15 | 7 | 11 | -3,5 | 0 | S31 | -20 | -3 | 2 | 4,75 |
| S16 | 5 | 13 | -1,5 | -1,5 | *Moy.* | *6,84* | *6,77* | *1,65* | *1,88* |

To describe these results, we executed a t-test analysis to determine if the differences were significant. The Table 6 presents the values obtained.

Table 6 – T-test results

| t-test (p-Value = 0,005) | | | |
|---|---|---|---|
| *Fluidity* | *Flexibility* | *Originality* | *Usefulness* |
| 0,22 | 1,13 | -1,64 | -0,13 |

To better understand the quantitative results, we asked the participants to write their personal impressions of the test on a web forum, anonymously. The main findings are presented in the next section.

### 5.2.2 Participants' view of the CEDA

Three main considerations emerged from the comments gathered in the web forum. First of all, the CEDA seems relevant and appropriate to assess participants' creativity. Comments noticeably stated that participants had the feeling that the test was pertinent and related to their creativity. Secondly, the CEDA seems to induce the "fixation" phenomenon explained by Ward [36, 37], in which the participants are influenced by memories of solutions (or examples) previously seen. Indeed, participants described specifically their struggle with finding other ideas than the ones they found the first time. Finally, the CEDA doesn't seem to motivate participants to think creatively. They find the test dull because the shapes combined with the given problems are not "exciting" and they felt the pressure to be creative even though the situation is not stimulant. Being a major factor in creative behavior [38, 39], motivation is an essential part to be considered in an assessment.

## 5.3 Qualitative results

To go deeper in our understanding of the course's impacts on students and go further than the quantitative data, we gathered data on what they thought about the new course. Using a web forum, students could voluntarily and anonymously, answer three questions on their academic experience. The principal results are presented in the next section.

### 5.3.1 Survey

Because we suspected that 13 weeks would be short to produce a quantitative positive result, we chose to use a mixed method research design and also collect qualitative data. At the end of the semester, an anonymous web survey was available for students to answer three questions. A majority of students (n = 18) gave pertinent and detailed responses.

The first question was "How relevant is a course like IND8108 in an engineering curriculum? Why?". Data analysis showed that the course is pertinent and that it is appropriate to have access to such a course in an engineering educational curriculum. Also, data revealed specific learnings students made during the course, like "the course allowed me to learn new tools to resolve problems", "the course gave me a better understanding of the creative process which could be relevant for engineers", "the course allowed me to learn another aspect of engineering", "I learned we had to be patient to be creative, we can practice and become better at it". They also thought that concepts learned during the semester could be applied in other courses like integration projects, "some techniques could of been directly applied in mechanical or industrial projects".

To the second question, that is "Explain if (and how) your perception of creativity has changed since the course's beginning?", data suggested that the students experienced  a change of perspective towards the complexity of creativity and its possible development. Students said that they now understand that creativity is a multifaceted field, that you can learn to become more creative (not born with it) and most importantly that it needs to be practiced. They also said that they learned to pay attention to their own cognitive creative process (meta-cognition) in order to better understand how they "work" during a creative activity.

The last question "Do you think you will be able incorporated the knowledge and skills you acquired during the course in a professional career?" brought data about the different approaches they could use to enhance their creativity and about their increase in self-confidence towards creative projects ("I learned numerous approaches to enhance creativity I could use", "I feel more confident about being open-minded" about creativity", "I will be more incline to participate in creative session, even as a leader").

## 6    DISCUSSION AND CONCLUSION

We succeeded in recruiting graduate and undergraduate students from different programs of a world-wide recognized school of engineering for an elective course on creativity. The students enlisted because they taught creativity is essential to future engineers. We conclude there is a need for the subject from the students' perspective.

Even if the CEDA was built from theory and a well-validated test on creativity, we chose to use it for this study. Still, there were no available published results from the author of the test and there were no before and after experimentation neither. We decided to use the test because of its theoretical validity. Although our results did not show an important increase in creativity (especially for originality) in the students who attended the course, the t-test produced still indicates a significant increase for fluidity, flexibility and usefulness. With no other research on the topic, we can't compare our findings to any other. Developing creativity in order that it shows in such a test takes probably more time than 12 weeks. Moreover, using the test twice brought up the fixation effect. The results suggest a positive influence on the students' creative performance but more research needs to done to improve the CEDA instrument and even better control the assessment.

However, our qualitative data showed that the students appreciated the course, its pertinence and its activities. Many students thought they were "more aware" of their creative process and acknowledged that they would use their learnings elsewhere in their program and as engineers. This allows us to pursue to offer the course. Also it is encouraging to keep documenting the effect of the course and of the different educative strategies used in the course in order to deepen our understanding of the cognitive processes that underline the creative process.

Combining the quantitative results that seems to significant improvement of creativity (three criteria out of four) from the beginning to the end of the course and the qualitative results that shows a positive experience from the students perspective, we conclude that an increase of creativity is possible to obtain with a course based on the development of cognitive process related to creativity. We found good elements and also opportunities for refinement.

Innovation emerges from creativity. Years of research were not enough to generate robust creativity programs capable of addressing major aspect of productivity and competitiveness. This resulted in an extremely wide and diversified offer in the field. This diversity reflects the lack of systematization of all the information collected.

The course started conservatively as an option in the "*Innovation technologique*" orientation. However, a more ambitious training program could also be considered. For these cognitive abilities to stay anchored and useable by the learners, continuous practise is important [4, 9]. We suggest two creativity trainings (*e.g.*, autumn of $1^{st}$ and $3^{rd}$ year) completed with the implication of creativity educators in the integrative projects (spring of every year). That will allow great opportunities to master the cognitive abilities and apply them to engineering projects.

Important steps have been made but more precise and validated elements need to be defined, especially concerning creativity training. We suggest that engineering faculties would better equip their students for the demands that will be placed upon them in their careers by teaching students how the creative process works, and how creativity can be developed.

## REFERENCES

[1]    R. K. Sawyer, "Educating for Innovation," *Thinking Skills and Creativity,* vol. 1, pp. 41-48, 2006.

[2]    H. Chesbrough, "The era of open innovation," *MIT Sloan Management Review,* pp. 35-41, 2003.

[3]    CNUCED. (2008). *Rapport sur l'économie créative 2008*.

[4]    R. K. Sawyer, *The Science of Human Innovation : Explaining Creativity*, 2nd ed. USA: Oxford University Press, 2012.

[5]    H. W. Chesbrough and M. M. Appelyard, "Open innovation and strategy," *California Management Review,* vol. 50, pp. 57-76, 2007.

[6]    D. Gantchev, "Assessing the economic contribution of creative industries: WIPO's experience," presented at the WIPO International Conference on IP and the Creative Industries, Geneva, Switzerland, 2007.

[7]    European Union. (2009, March 1st, 2015). *Manifesto*. Available: http://www.create2009.europa.eu/fileadmin/Content/Downloads/PDF/Manifesto/manifesto.en.pdf

[8]    Government of Canada. (2009). "Canada at the Leading Edge : Common Vision, Concerted Plan." Retrieved from http://www.nserc-crsng.gc.ca/_doc/Reports-Rapports/SSHRC-NSERC-CIHR_Final_eng.pdf.

[9]    R. K. Sawyer, *Zig Zag : the surprising path to greater creativity*, 1st ed. USA: Jossey-Bass, 2013.

[10]   B. Nussbaum, *Creative Intelligence : Harnessing the Poer to Create, Connect, and Inspire*. USA: Harper Business, 2013.

[11] S. Zappe, T. Litzinger, and S. Hunter, "Integrating the Creative Process into Engineering Courses: Description and Assessment of a Faculty Workshop.," in *Annual Conference of the American Society for Engineering Education*, San Antonio, TX, 2012.

[12] C. Zhou, "Integrating creativity training into Problem and Project- Based Learning curriculum in engineering education," *European J. of Engin. Education,* vol. 37, pp. 488-499, 2012.

[13] C. Baillie and P. Walker, "Fostering Creative Thinking in Student Engineers," *European J. of Engin. Education,* vol. 23, pp. 35-44, 1998.

[14] G. Scott, L. E. Leritz, and M. D. Mumford, "The effectiveness of creativity training: A quantitative review," *Creativity Research Journal,* vol. 16, pp. 361-388, 2004.

[15] G. Scott, L. E. Leritz, and M. D. Mumford, "Types of creativity training: Approaches and their effectiveness," *The Journal of Creative Behavior,* vol. 38, pp. 149-179, 2004.

[16] M. Murdock and S. Keller-Mathers, "Programs and Courses in Creativity," in *Encyclopedia of Creativity.* vol. 2, M. Runco and S. Pritzker, Eds.,  USA: Academic Press, 2011, pp. 266-270.

[17] J. E. Mills and D. F. Treagust. (2003). *Engineering Education – Is problem-based or project based Learning the answer ?*

[18] W. B. Stouffer, J. S. Russell, and M. G. Oliva, "Making The Strange Familiar: Creativity and the Future of Engineering Education," in *Ame. Society for Engin. Education Annual Conf. & Exposition*, 2004.

[19] R. J. Sternberg, T. I. Lubart, J. C. Kaufman, and J. E. Pretz, "Creativity," in *Cambridge handbook of thinking and reasoning*, K. J. Holyoak and R. G. Morrison, Eds.,  Cambridge: Cambridge University Press, 2005, pp. 351-370.

[20] M. J. Wilkenfeld and T. B. Ward, "Similarity and Emergence in Conceptual Combination," *J. of Memory and Language,* vol. 45, pp. 21-38, 2001.

[21] M. Lande, "Student Engineers Learning to Become Designers," in *Seventh ACM Conference on Creativity and Cognition*, Berkeley, Californie, 2009.

[22] L. Pappano. (2014, Feb. 10th, 2014). *Learning to Think Outside the Box, Creativity Becomes an Academic Discipline*. Available: http://www.nytimes.com/2014/02/09/education/edlife/creativity-becomes-an-academic-discipline.html

[23] T. Lubart, "Models of the Creative Process: Past, Present and Future," *Creativity Research Journal,* vol. 13, pp. 295-308, 2010.

[24] Z. E. Liu and D. J. Schönwetter, "Teaching Creativity in Engineering," *International Journal of Engineering Education,* vol. 20, pp. 801-808, 2004.

[25] J. H. Newton and K. S. McGrew, "Introduction to the Special Issu : Current research in Cattel-Horn-Carroll- Based Assessment," *Psychology in the Schools,* vol. 4, pp. 621-634, 2010.

[26] L. Gabora, "Revenge of the 'neurds': Characterizing creative thought in terms of the structure and dynamics of human memory," *Creativity Research Journal,* vol. 22, pp. 1-13, 2010.

[27] M. Benedek, T. Könen, and A. C. Neubauer, "Associative Abilities Underlying Creativity," *Psychology of Aesthetics, Creativity and the Arts,* vol. 6, pp. 273-281, 2012.

[28] M. Batey and A. Furnham, "creativity, Intelligence, and Personality : A Critical Review of the Scattered Literature," *Genetic, Social, and General Psychology Monographs,* vol. 132, pp. 355-429, 2006.

[29] M. J. Avitia and J. C. Kaufman, "Beyond g and c: The Relationship of Rated Creativity to Long-Term Storage and Retrieval (Glr)," *Psychology of Aesthetics, Creativity and the Arts,* vol. 8, pp. 293-302, 2014.

[30] C. Charyton, *Creative Engineering Design Assessment: Background, Directions, Manual, Scoring Guide and Uses*: Springer London, 2014.

[31] D. Harris, "The Development and Validation of a Test of Creativity in Engineering," *Journal of Applied Psychology,* vol. 44, pp. 254-257, 1960.

[32]     D. Harris, "The Development and Validation of a Creativity Test in Engineering," *Journal of Applied Psychology,* vol. 44, pp. 254-257, 1960.

[33]     G. Lemons, "Diverse Perspectives of Creativity Testing Controversial Issues When Used for Inclusion Into Gifted Programs," *Journal for the Education of the Gifted,* vol. 34, pp. 742-772, 2011.

[34]     D. Piffer, "Can creativity be measured? An attempt to clarify the notion of creativity and general directions for future research," *Thinking Skills and Creativity,* vol. 7, pp. 258-264, 2012.

[35]     C. Charyton and J. A. Merrill, "Assessing General Creativity and Creative Engineering Design in First Year Engineering Students," *J. of Engineering Education,* pp. 145-156, 2009.

[36]     T. B. Ward, "Creative cognition as a window on creativity," *Methods,* vol. 42, pp. 28-37, 2007.

[37]     T. B. Ward, S. M. Smith, and R. A. Finke, "Creative Cognition," in *Handbook of Creativity*, R. J. Sternberg, Ed.,  Cambridge, UK: Cambridge Univ. Press, 1999, pp. 189-212.

[38]     R. J. Sternberg, "The Nature of Creativity," *Creativity Research Journal,* vol. 18, pp. 87-98, 2006.

[39]     M. Barack and N. Goffer, "Fostering Systematic Innovative Thinking and Problem Solving: Lessons Education Can Learn From Industry," *International Journal of  Technology and design Education,* vol. 12, pp. 227-247, 2002.